\documentclass{PoS}

\title{Baryon masses with dynamical twisted mass fermions}

\ShortTitle{Baryon masses with dynamical twisted mass fermions}

\author{\speaker{Constantia Alexandrou}
         \thanks{On behalf of the European Twisted Mass Collaboration},   
Tomasz Korzec, Giannis Koutsou\\
  Department of Physics, University of Cyprus, P.O. Box 20537,
 1678 Nicosia, Cyprus\\
        E-mail: \email{alexand@ucy.ac.cy, korzec@ucy.ac.cy, koutsou@ucy.ac.cy}}
\author{Remi Baron, Pierre Guichon\\
 DAPNIA/SPhN B\^at 703, Orme des Merisiers, CEA Saclay, France}
\author{Mariane Brinet, Jaume Carbonell, Vincent Drach\\
Laboratoire de Physique Subatomique et Cosmologie,
               53 avenue des Martyrs, 38026 Grenoble, France\\
Email: \email{mariane@lpsc.in2p3.fr}}
\author{Zhaofeng Liu, Olivier P\`ene\\
Laboratoire de Physique Th\'eorique,
               UMR8627, Universit\'e Paris XI, 91405 Orsay-Cedex, France\\
Email: \email{Olivier.Pene@th.u-psud.fr}}
\author{Carsten Urbach\\
Theoretical Physics Division, Dept. of Mathematical Sciences, 
University of Liverpool, Liverpool L69 7ZL, UK\\
Email: \email{Carsten.Urbach@physik.hu-berlin.de}}

\abstract{We present results on the mass of the nucleon and  
the $\Delta$ using two
 dynamical degenerate twisted mass quarks. The evaluation
is performed at four quark masses corresponding to a pion mass in the range
of 690-300~MeV on lattices of size 2.1~fm and 2.7~fm.
We check for cutoff effects by
evaluating these baryon masses on lattices of spatial size 2.1~fm with
lattice spacings $a(\beta=3.9)=0.0855(6)$~fm 
and $a(\beta=4.05)=0.0666(6)$~fm, 
determined from the pion sector and find them to
be within our statistical errors.
Lattice results
are extrapolated to the physical limit using continuum chiral perturbation
theory. 
 The nucleon mass at the physical point provides a determination
of the lattice spacing. Using heavy baryon chiral perturbation theory
at ${\cal O}(p^3)$ we find $a(\beta=3.9)=0.0879(12)$~fm, with a
 systematic error  due to the chiral
extrapolation estimated to be about the same as the statistical error. This
value of the lattice spacing is in
good agreement 
with the value  determined from the pion sector. 
We check for isospin breaking in the $\Delta$-system. We find that
$\Delta^{++,-}$ and $\Delta^{+,0}$ are almost degenerate pointing to
 small flavor violating effects.}

\FullConference{The XXV International Symposium on Lattice Field Theory\\
		 July 30 - August 4 2007\\
		 Regensburg, Germany}

\usepackage{graphicx}
\begin{document}

\newcommand{\be}{\begin{equation}}
\newcommand{\ee}{\end{equation}}
\newcommand{\beq}{\begin{eqnarray}}
\newcommand{\eeq}{\end{eqnarray}}

\section{Introduction}
Twisted mass fermions provide a promising formulation of lattice QCD that
allows for automatic ${\cal O}(a)$ improvement, infrared regularization 
of small
eigenvalues and fast dynamical simulations~\cite{TM intro}.
We use the tree-level Symanzik improved gauge action and work
 at maximal twist to realize ${\cal O}(a)$-improvement.
Recent results obtained in the pion sector 
give an accurate evaluation of the
low energy constants $\bar{l}_3$ and $\bar{l}_4$~\cite{fpi, Urbach}, which
lead to the most accurate determination of the S-wave $\pi\pi$ scattering 
lengths~\cite{Leutwyler}. In this work we study the light  baryon sector.

The fermionic action for two degenerate flavors of quarks
 in twisted mass QCD is given by
\be
S_F= a^4\sum_x  \bar{\psi}(x)\bigl(D_W[U] + m_0 
+ i \mu \gamma_5\tau^3  \bigr ) \psi(x)
\label{S_tm}
\ee
with $D_W[U]$ the massless Dirac operator, $m_0$ the bare
untwisted quark mass and $\mu$ the bare twisted mass. 
The twisted mass term in the fermion action of Eq.~(\ref{S_tm})
 breaks isospin symmetry since the u- and d-quarks differ
 by having opposite signs 
for the  $\mu$-term. This isospin breaking is a 
cutoff effect of ${\cal O}(a^2)$. However
the up- and down-propagators satisfy
  $G_u(x,y) = \gamma_5 G_d^\dagger(y,x)\gamma_5$, 
which  means that  
two-point correlators are equal with their
hermitian conjugate with u- and d-quarks interchanged.
Since the  masses are computed from real correlators this leads
to the following pairs being
degenerate:
$\pi^+$ and $\pi^-$, proton and
neutron and $\Delta^{++}(\Delta^+)$ and $\Delta^{-}(\Delta^0)$.
 A theoretical analysis~\cite{Rossi} shows
that potentially large ${\cal O}(a^2)$ effects that appear in the $\pi^0$-mass
are suppressed in all other quantities.
 Calculation of the mass of $\pi^0$, 
which  requires the evaluation of disconnected
diagrams, has been carried out  confirming large ${\cal O}(a^2)$-effects.
In the baryon sector we can study isospin breaking 
 by evaluating the mass difference
between $\Delta^{++}(\Delta^-)$ and $\Delta^{+}(\Delta^0)$.
Since no disconnected contributions enter we can obtain an
accurate evaluation of isospin splitting and its dependence on 
the lattice spacing.  

\section{Lattice techniques}
The parameters of the calculation are collected in Table~\ref{Table:params}.
They
span a pion mass range from 300-690~MeV. At a pion mass of about 300~MeV
we have simulations for lattices of
 spatial size, $L_s=2.1$~fm and $L_s=2.7$~fm at $\beta=3.9$ allowing to
check finite size effects. We provide a first check of 
finite $a$-effects by comparing results at $\beta=3.9$ and $\beta=4.05$. 

The masses of the nucleon and the $\Delta$'s are extracted from two-point
correlators using the
standard interpolating fields, which for the proton, 
the $\Delta^{++}$ and $\Delta^{+}$, are given by
\beq \nonumber
J_p &=& \epsilon_{abc}\bigl( u^T_a C\gamma_5 d_b\bigr)u_c, \hspace*{1cm}
J_{\Delta^{++}} = \epsilon_{abc}\bigl( u^T_a C\gamma_\mu u_b\bigr)u_c \\
J_{\Delta^{+}} &=& \frac{1}{\sqrt{3}}\epsilon_{abc}\biggl[
2\bigl( u^T_a C\gamma_\mu d_b\bigr)u_c +\bigl( u^T_a C\gamma_\mu u_b\bigr)d_c
\biggl] \quad.
\label{interpolate}
\eeq

\begin{table} 
\begin{center}
\begin{tabular}{c|ccccc}
\hline
\multicolumn{6}{c}{$\beta=3.9$, 
$a=0.0855(6)$~fm from $f_\pi$~\cite{Urbach}}\\
\hline 
 $24^3\times 48$, $L_s=2.1$~fm &$\mu$         & 0.0040      &   0.0064     &  0.0085     &   0.010 \\ 
 &$m_\pi$~(GeV) & 0.3131(16) & 0.3903(9) & 0.4470(12) & 0.4839(12)\\
$32^3\times 64$, $L_s=2.7$~fm &$\mu$ & 0.004 & & & \\
&  $m_\pi$~(GeV) & 0.3082(55) & & & 
\\\hline
\hline
%\multicolumn{5}{c}{$32^3\times 64$, $L=2.7$~fm}\\ \hline
% $\mu$ & 0.004 & & & \\
% $m_\pi$~(GeV) & 0.3082(55) & & & 
%\\\hline
\multicolumn{6}{c}{ $\beta=4.05$, $a=0.0666(6)$~fm from $f_\pi$~\cite{Urbach}}\\
\hline
$32^3\times 64$, $L_s=2.1$~fm &$\mu$         & 0.0030     & 0.0060     & 0.0080     & 0.010\\
 &$m_\pi$~(GeV) & 0.3070(18) & 0.4236(18) & 0.4884(15) & 0.6881(18) \\
\hline
\end{tabular}
\caption{The parameters of our calculation.}
\label{Table:params}
\end{center}
\vspace*{-0.8cm}
\end{table}

\begin{figure}[ht]
\vspace*{-0.3cm}
\begin{minipage}{6.cm}\vspace*{-1.3cm}
\hspace*{-0.8cm}
\ifpdf
{\mbox{\includegraphics[height=6cm,width=7.5cm]{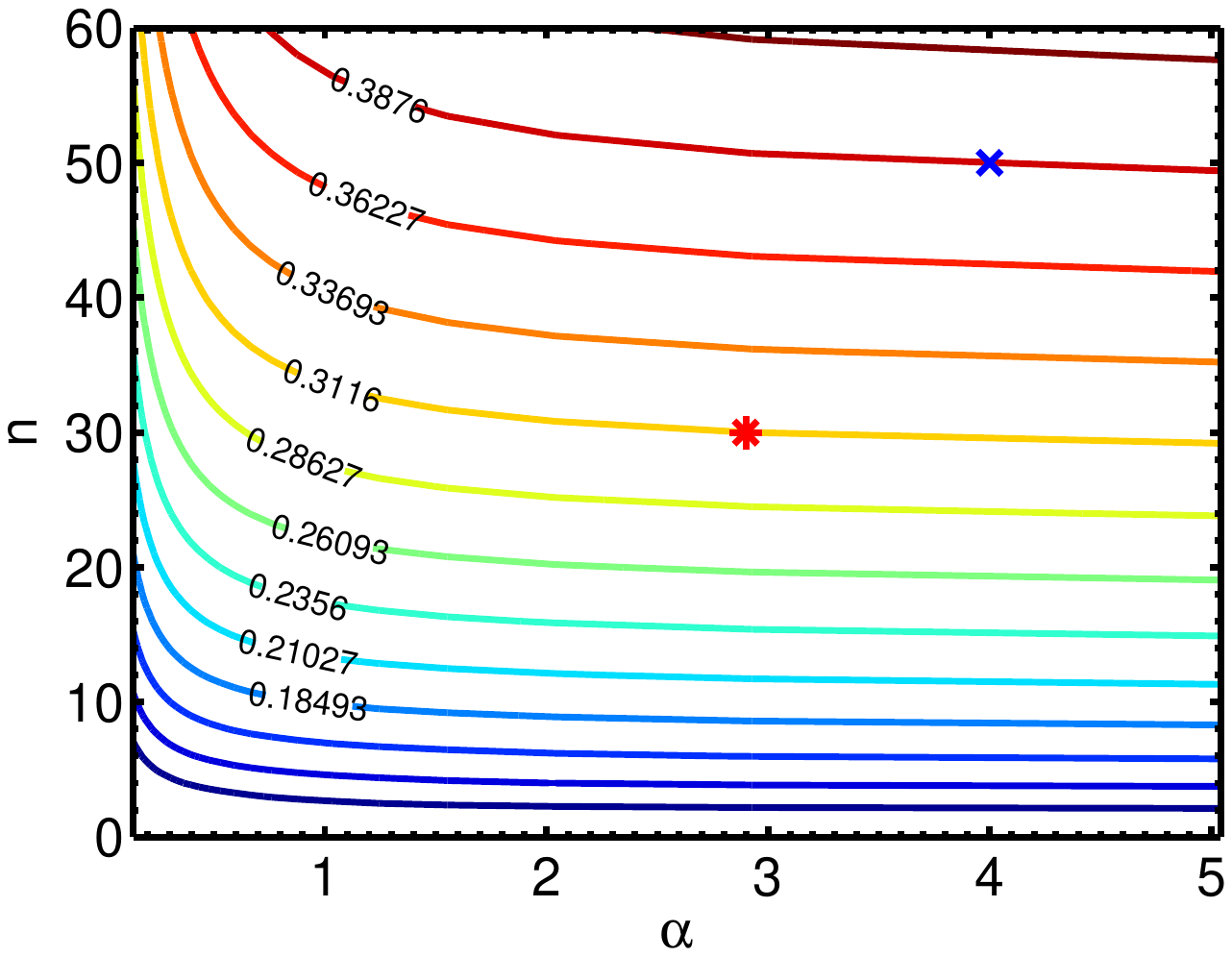}}}
\else
{\mbox{\includegraphics[height=6cm,width=7.5cm]{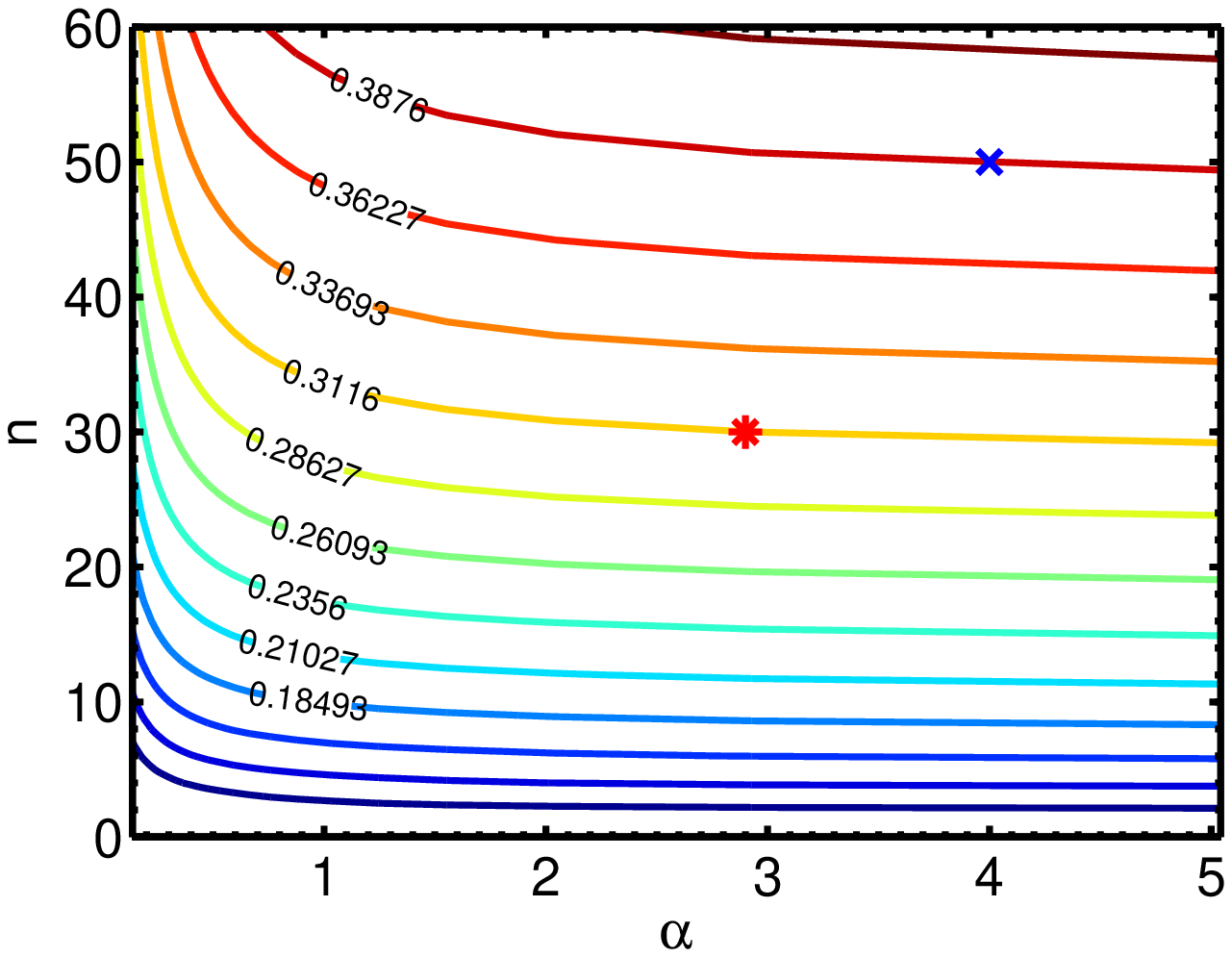}}}
\fi
\vspace*{-0.5cm}
\caption{Lines of constant r.m.s radius as function of the smearing
parameters $\alpha$ and $n$. The asterisk shows the 
values  $\alpha=2.9$, $n=30$ and the
cross  $\alpha=4.0$, $n=50$. }
\label{fig:contour}
\end{minipage}\hspace*{0.5cm}
\begin{minipage}{8.5cm}
\ifpdf
{\mbox{\includegraphics[height=5.5cm,width=8cm]{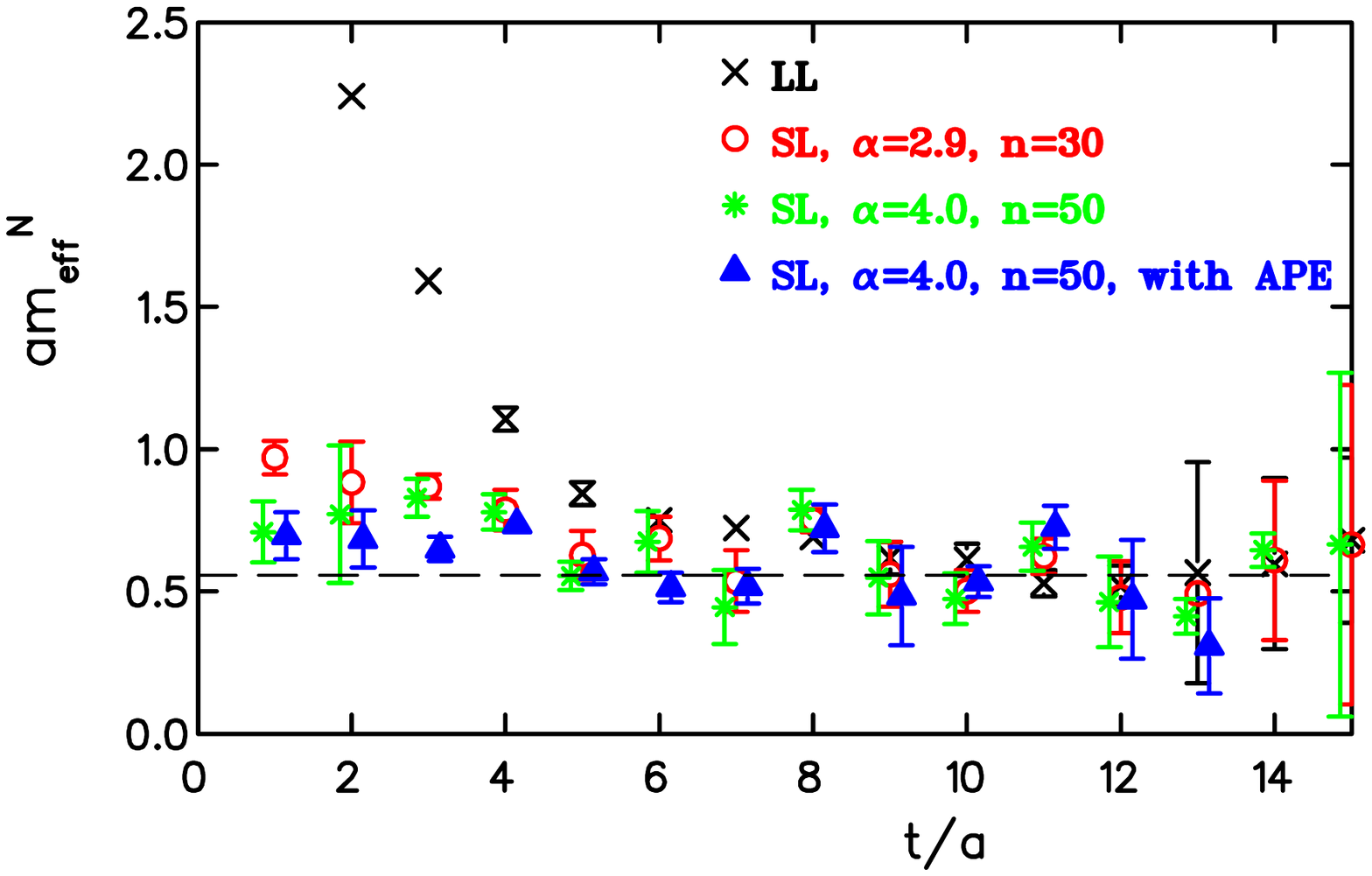}}}
\else
{\mbox{\includegraphics[height=5.5cm,width=8cm]{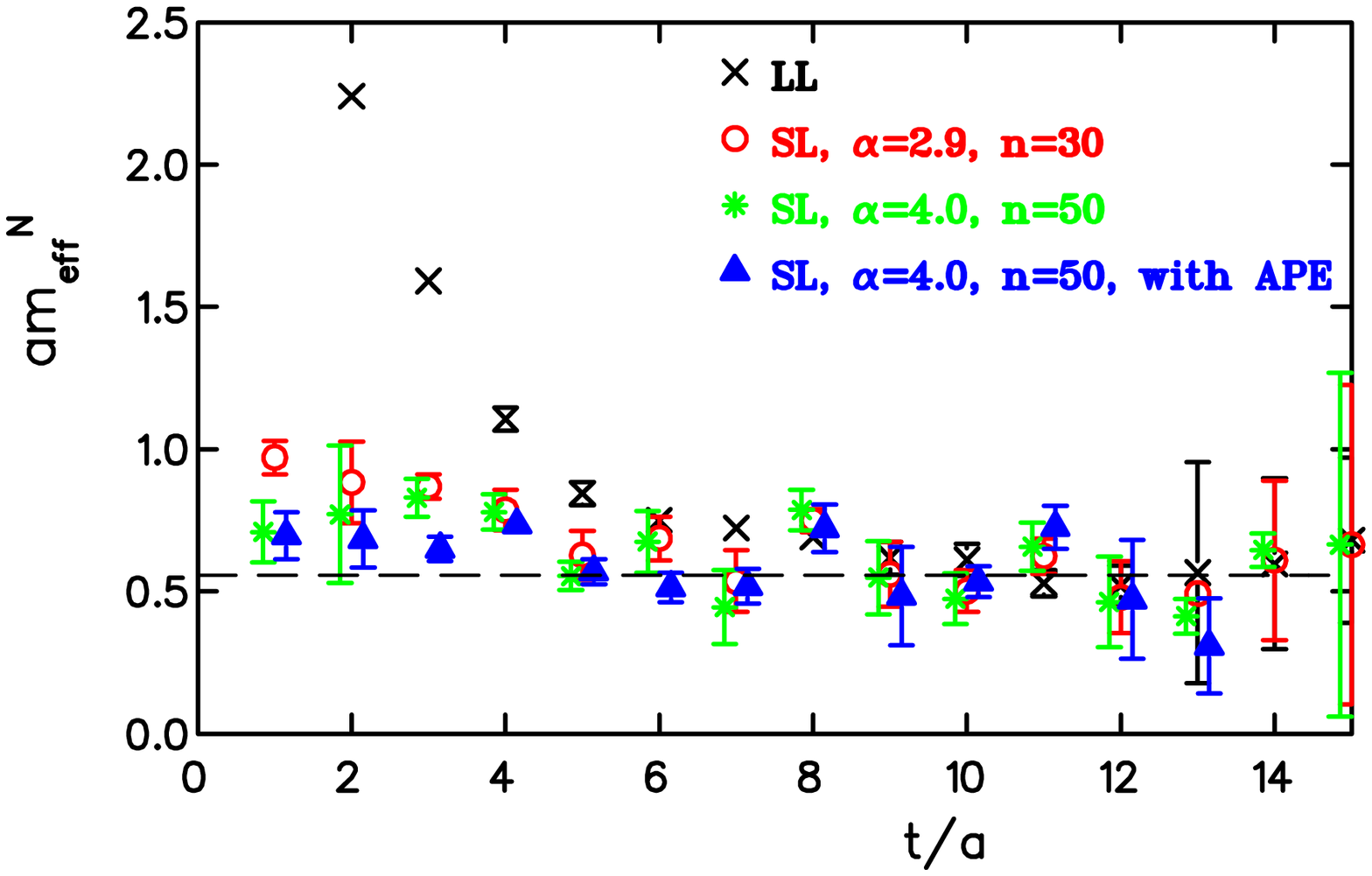}}}
\fi
\caption{$m_{\rm eff}^N$ versus time separation both in
lattice units. Crosses show
results using local sink and source (LL), 
circles (asterisks) using Gaussian smearing at the sink (SL) with
 $\alpha=2.9$ and $n=30$ ( $\alpha=4$ and $n=50$),
  and filled triangles with $\alpha=4$ and $n=50$ 
 and APE smearing. The dashed line is the plateau value when APE smearing
is used.}
\label{fig:meff_opt}
\end{minipage}
\end{figure}

\noindent
 Local interpolating fields %of Eq.~(\ref{interpolate})
 are not
optimal for suppressing excited state contributions. We instead apply
 Gaussian
 smearing to each  quark field,  $q({\bf x},t)$:
$q^{\rm smear}({\bf x},t) = \sum_{\bf y} F({\bf x},{\bf y};U(t)) q({\bf y},t)$
using the gauge invariant smearing function  
\be 
F({\bf x},{\bf y};U(t)) = (1+\alpha H)^ n({\bf x},{\bf y};U(t)),
\ee
constructed from the
hopping matrix,
$
H({\bf x},{\bf y};U(t))= \sum_{i=1}^3 \biggl( U_i({\bf x},t)\delta_{{\bf x,y}-i} +  U_i^\dagger({\bf x}-i,t)\delta_{{\bf x,y}+i}\biggr).
$
The parameters $\alpha$ and $n$ are varied so 
that the root mean square (r.m.s)  radius  obtained
 using the proton interpolating field is in the range of 0.3-0.4~fm.
In Fig.~\ref{fig:contour} we show lines of constant r.m.s radius as
we vary $\alpha$ and $n$.
The larger the  $n$ the more time consuming is the smearing procedure.
 On the other hand,
for  $\alpha\stackrel{>}{\sim}1$, increasing further $\alpha$ 
does not reduce $n$ significantly. 
Therefore, we choose a value of $\alpha$ large enough 
so that the weak $n$-dependence sets in, and we
adjust $n$ to obtain the required value of the r.m.s radius.
We consider two sets for these parameters  
giving r.m.s radius 0.31~fm and 0.39~fm, as  
shown in Fig.~\ref{fig:contour}.
In Fig.~\ref{fig:meff_opt}, we show the nucleon effective mass,
 $m_{\rm eff}^N=-\log(C(t)/C(t-1))$ with $C(t)$ the
nucleon correlator, 
  for 10 configurations  at $\beta=3.9$
and $\mu=0.0085$. For the optimization of the parameters 
we apply Gaussian smearing at the
source, whereas for  the sink we use local interpolating fields
so that no additional inversions are needed when we change $\alpha$ and $n$.
As can be seen, for both sets of smearing parameters,
the excited state contributions are suppressed with the set $\alpha=4$, $n=50$
producing  a plateau a couple of time slices earlier.
 If, in addition, we apply APE smearing to the spatial links 
that enter the hopping matrix,
then  gauge noise  is reduced resulting in a better identification of the 
plateau.
Therefore for all computations at  $\beta=3.9$ we  use Gaussian smearing with 
$\alpha=4$ and $n=50$. We apply smearing at the source
and compute the mass using both local (LS) and smeared sink (SS).
For $\beta=4.05$ we readjust the parameters so that the nucleon 
r.m.s radius is still about 0.39~fm, obtaining  $\alpha=4$ and $n=70$.
In all cases we apply APE smearing to the gauge links that
are used in $F({\bf x},{\bf y};U(t))$.
We note that Gaussian smearing is very effective as compared to, for example,
  fuzzing on links joining quarks at different sites.

\begin{figure}[h]
\begin{minipage}{7.5cm}\hspace*{-0.5cm}\vspace*{-0.3cm}
\ifpdf
{\mbox{\includegraphics[height=7cm,width=7.5cm]{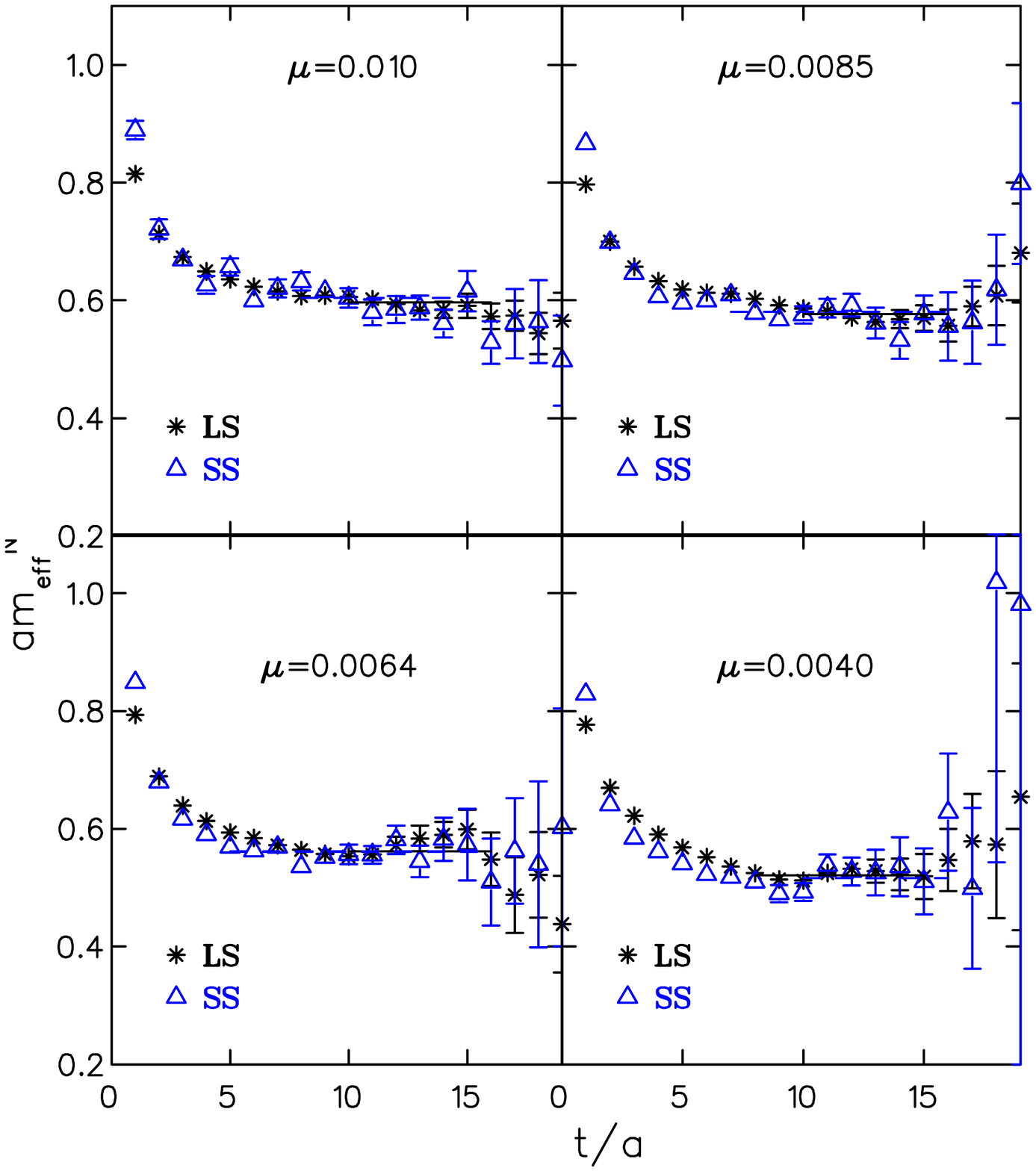}}}
\else
{\mbox{\includegraphics[height=7cm,width=7.5cm]{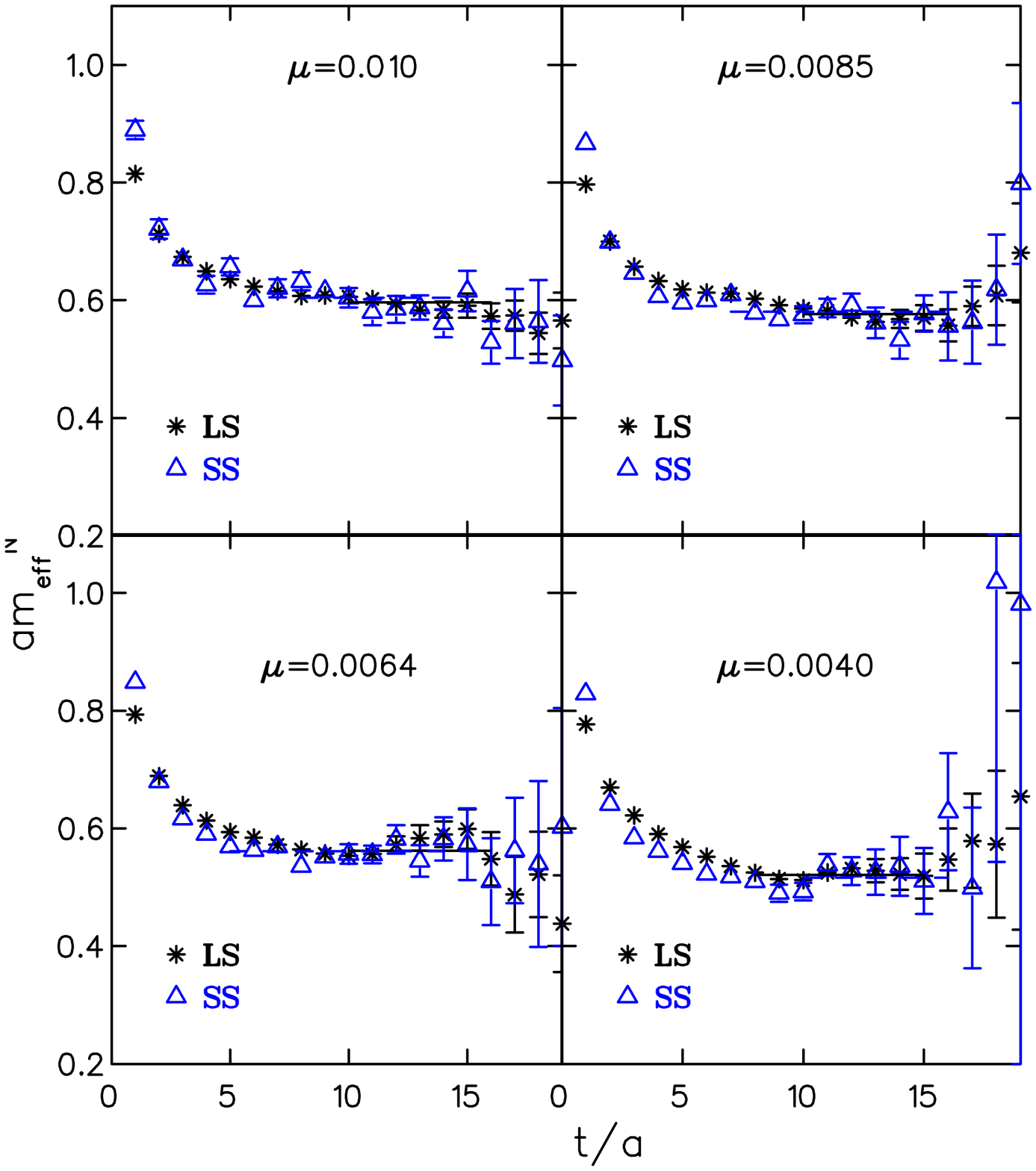}}}
\fi
\caption{Nucleon effective mass  (LS: asterisks, SS: open triangles)
for $\beta=3.9$ 
versus
time separation in lattice units. }
\label{fig:nucleon meff}
\end{minipage}\hspace*{0.3cm}
\begin{minipage}{7.5cm}\hspace*{-0.5cm}
\ifpdf
{\mbox{\includegraphics[height=7cm,width=7.5cm]{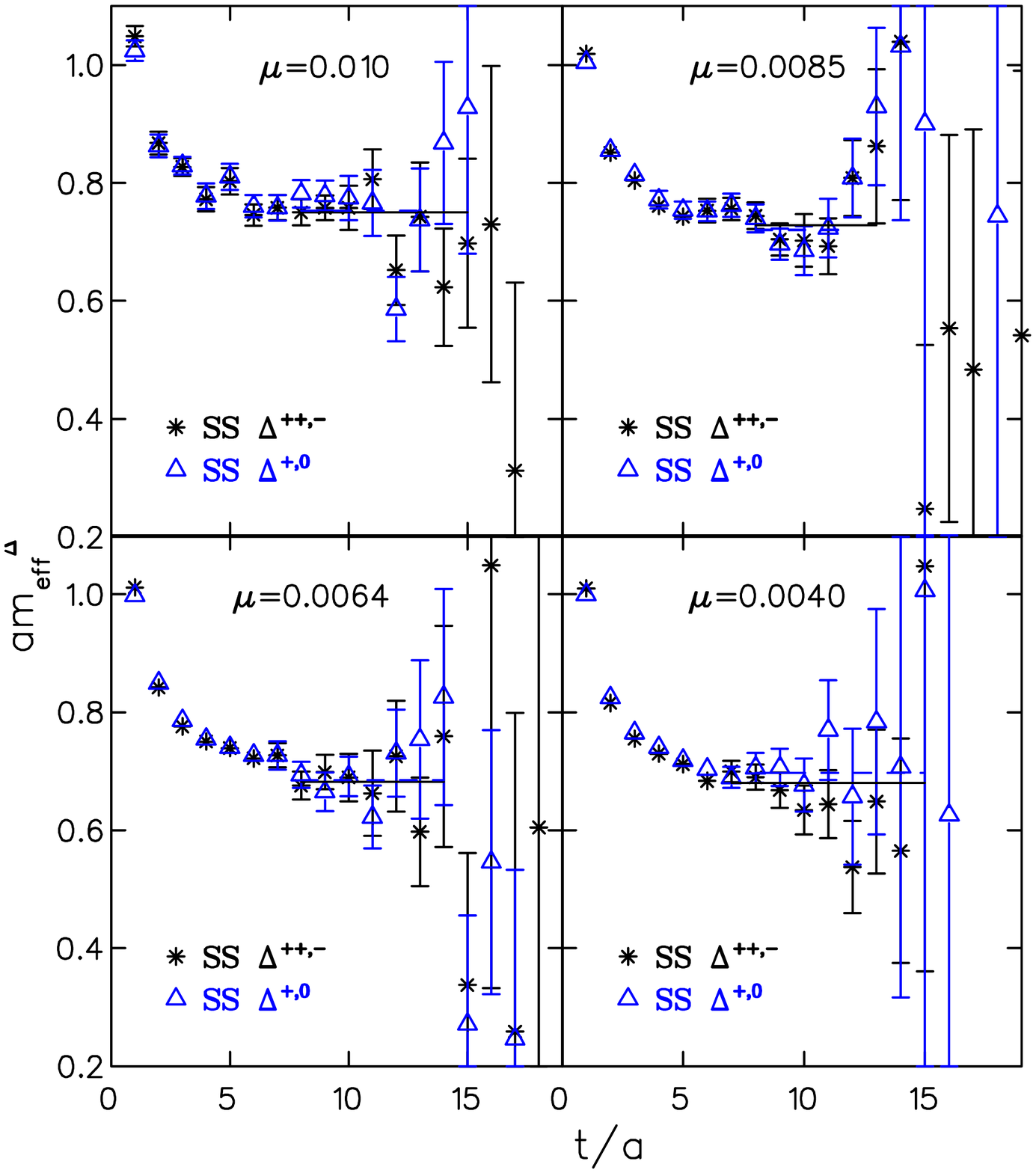}}}
\else
{\mbox{\includegraphics[height=7cm,width=7.5cm]{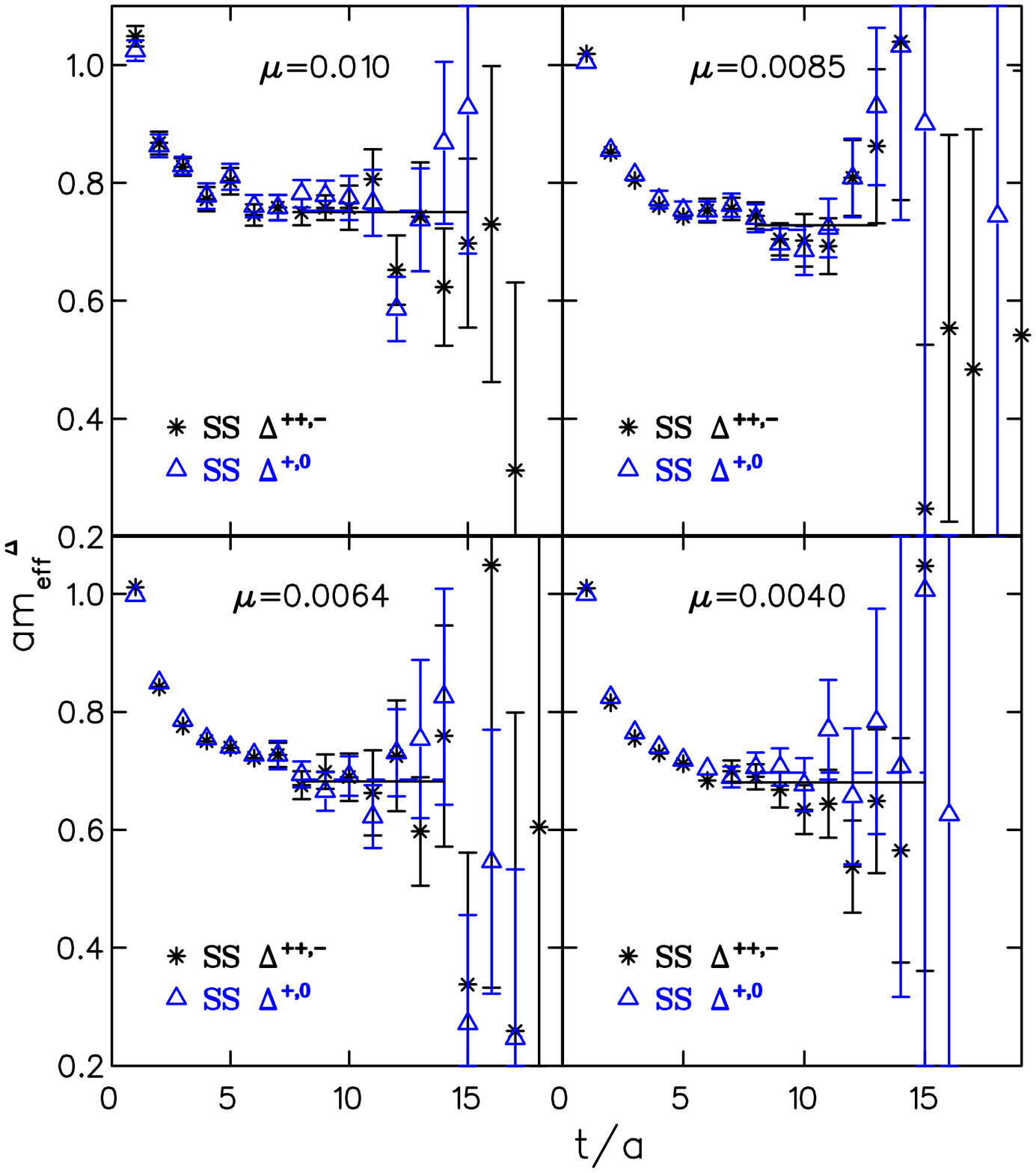}}}
\fi
\vspace*{-0.3cm}
\caption{$\Delta^{++,-}$ (asterisks) and $\Delta^{+,0}$ (open triangles)
 effective masses for $\beta=3.9$ %and 
%$\mu=0.010$, $0.0085$, $0.0064$ and $0.0040$
versus
time separation in lattice units. }
\label{fig:delta meff}
\end{minipage}
\vspace*{-0.3cm}
\end{figure}

 The nucleon 
effective masses  obtained 
using correlators with smeared source and local  or smeared  sink
 for the four 
 $\mu$-values at $\beta=3.9$ are shown in
Fig.~\ref{fig:nucleon meff}, where we average over the
proton and neutron correlators.
In Fig.~ \ref{fig:delta meff} we show,  for the same $\mu$-values, 
the $\Delta$ effective masses 
after averaging the correlators obtained using smeared source and sink
over the degenerate pairs $\Delta^{++}$,  $\Delta^-$ 
and  $\Delta^{+}$, $\Delta^0$ .
 As can be seen,
 the quality of the plateaus in the nucleon
case is better than in case of the $\Delta$. This explains why results
on the $\Delta$ mass have larger errors requiring more statistics 
for a  reliable 
 determination. The errors are evaluated using jackknife and
the $\Gamma$-method~\cite{Wollf} to check consistency.
In all the figures we show the errors obtained with the latter method.

\vspace*{-0.3cm}

\section{Results}
\begin{figure}[h]   
\begin{minipage}{7.4cm}
\ifpdf
{\mbox{\includegraphics[height=5.5cm,width=7.5cm]{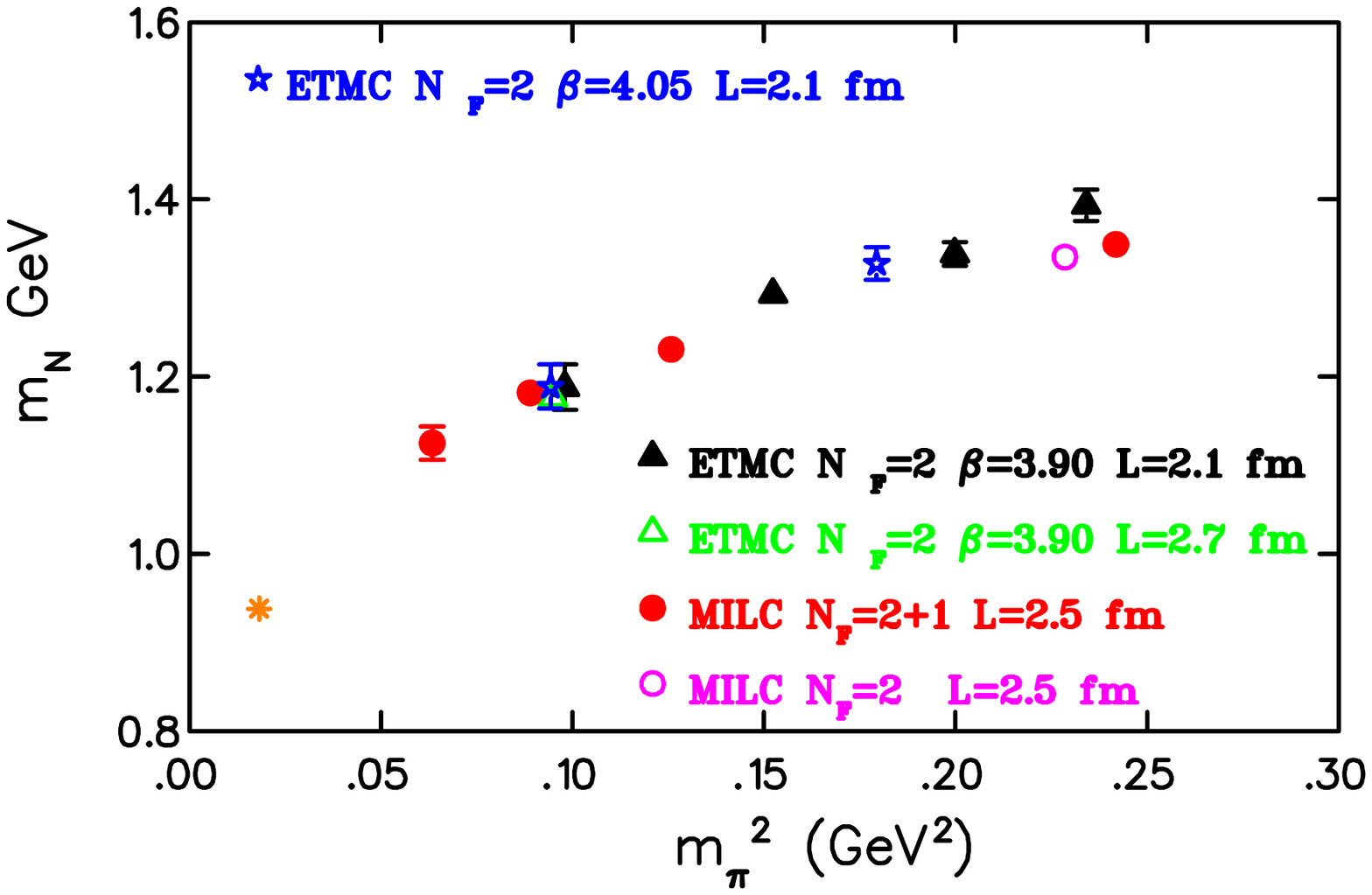}}}
\else
{\mbox{\includegraphics[height=5.5cm,width=7.5cm]{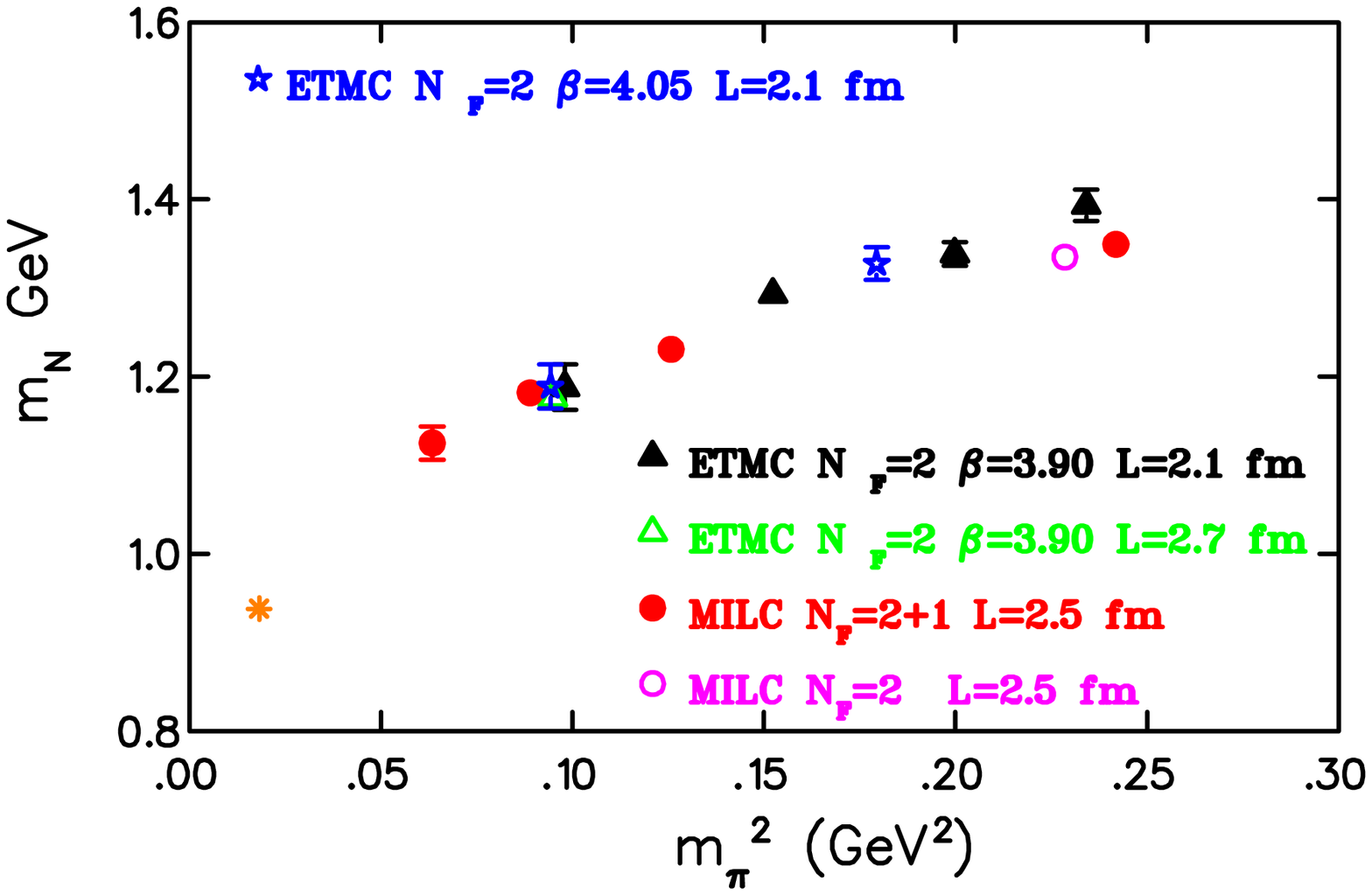}}}
\fi
\caption{The nucleon mass as a function of $m_\pi^2$
for $\beta=3.9$ on a lattice of size $24^3\times 48$ (filled triangles)
and on a lattice of size $32^3\times 64$ (open triangles). Results
at $\beta=4.05$ are shown with the stars. The physical nucleon
mass is shown with the asterisk.
Results with dynamical staggered
fermions for $N_F=2+1$ (filled circles) and $N_F=2$ (open circle)
on a lattice of size $20^3\times 64$ with $a=0.125$~fm 
are from Ref.~\cite{staggered}.} 
\label{fig:mn}
\end{minipage}\hspace*{0.3cm}
\begin{minipage}{7.7cm}\vspace*{-2.2cm}
\ifpdf
{\mbox{\includegraphics[height=5.5cm,width=7.5cm]{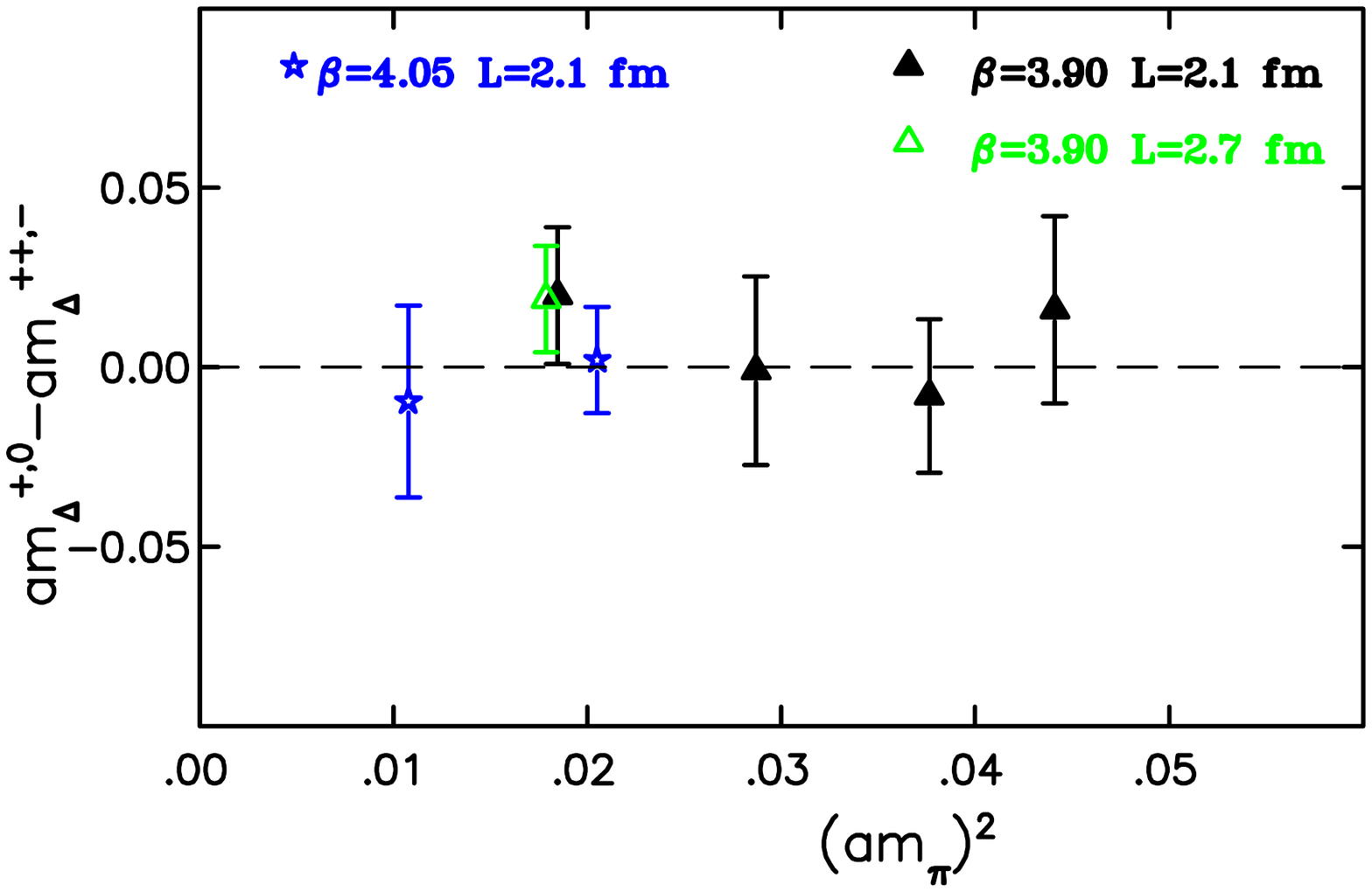}}}
\else
{\mbox{\includegraphics[height=5.5cm,width=7.5cm]{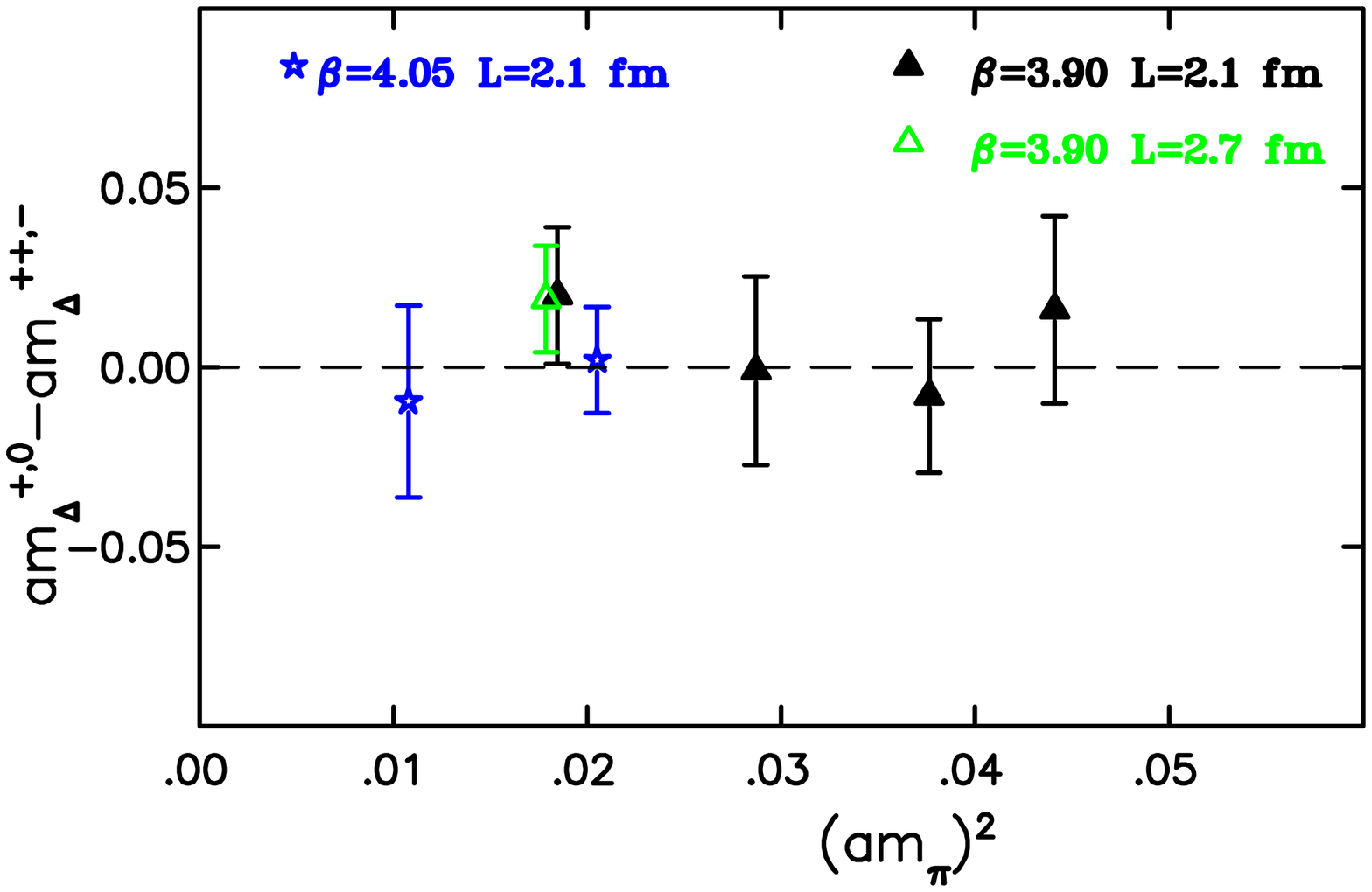}}}
\fi
\caption{The mass 
 splitting between $\Delta^{+,0}$ and $\Delta^{++,-}$
  as a function of $m_\pi^2$ both in lattice units. The notation is the
same as in Fig.~5.}
\label{fig:Dm}
\end{minipage}
\end{figure}

We show our results for the nucleon mass as 
a function of  $m_\pi^2$
in Fig.~\ref{fig:mn}, where we use $a=0.0855$ at
$\beta=3.9$ and $a=0.0666$ at $\beta=4.05$, determined from
$f_\pi$~\cite{fpi}, to convert lattice results
to physical units. As can be
seen, the results at these two $\beta$-values show good scaling 
pointing to small cutoff effects. For the three
larger pion masses  $m_\pi L_s \ge 4$, whereas for the smallest
value  $m_\pi(\mu=0.004) L_s \sim 3.2$. Applying the
resummed L\"uscher formula to the nucleon mass and using 
 the knowledge of the $\pi N$ scattering amplitude to ${\cal O}(p^2)$
and  ${\cal O}(p^4)$ it was shown that, for $L_s\sim 2$~fm and 
$m_\pi\sim 300$~MeV, the volume corrections are small 
 being estimated  to be 
about (3-5)\%~\cite{Colangelo}.
We calculate the nucleon mass increasing the spatial length
of the lattice from $2.1$~fm to $2.7$~fm
so that  $m_\pi(\mu=0.004) L_s\sim 4.3$. 
 If  $\Delta m_N \equiv m_N(L_s/a=24)-m_N(L_s/a=32)$ then
we find that $\Delta m_N/m_N(L_s/a=32)=0.01 \pm 0.02$ at our 
smallest quark mass, i.e. consistent with zero within our statistical error
but also within the estimated error range of Ref.~\cite{Colangelo}.
In Fig.~\ref{fig:mn} we include, for comparison,
 results obtained with dynamical 
staggered fermions from Ref.~\cite{staggered}.
As can be seen, the results using  these two formulations  
are consistent with each other.

In Fig.~\ref{fig:Dm} we show our results for the mass difference
between the averaged mass
of the pairs $\Delta^{++}$, $\Delta^-$ and  $\Delta^{+}$, 
$\Delta^{0}$. As can be seen, the splitting is consistent with
zero, indicating that isospin breaking in the $\Delta$ system is small.

\begin{figure}[h]
\begin{minipage}[h]{7.cm}
\ifpdf
{\mbox{\includegraphics[height=5.5cm,width=7.2cm]{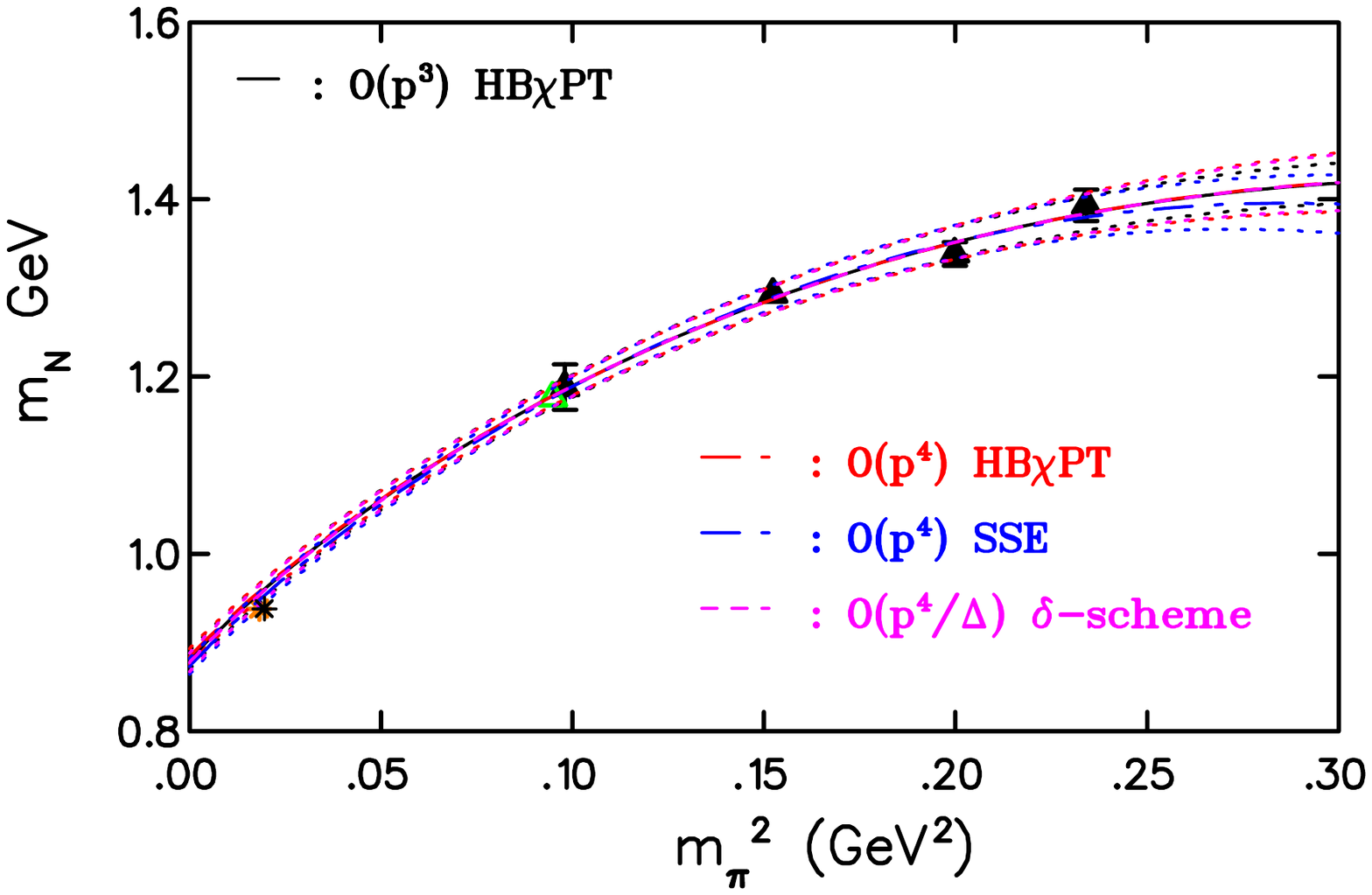}}}
\else
{\mbox{\includegraphics[height=5.5cm,width=7.2cm]{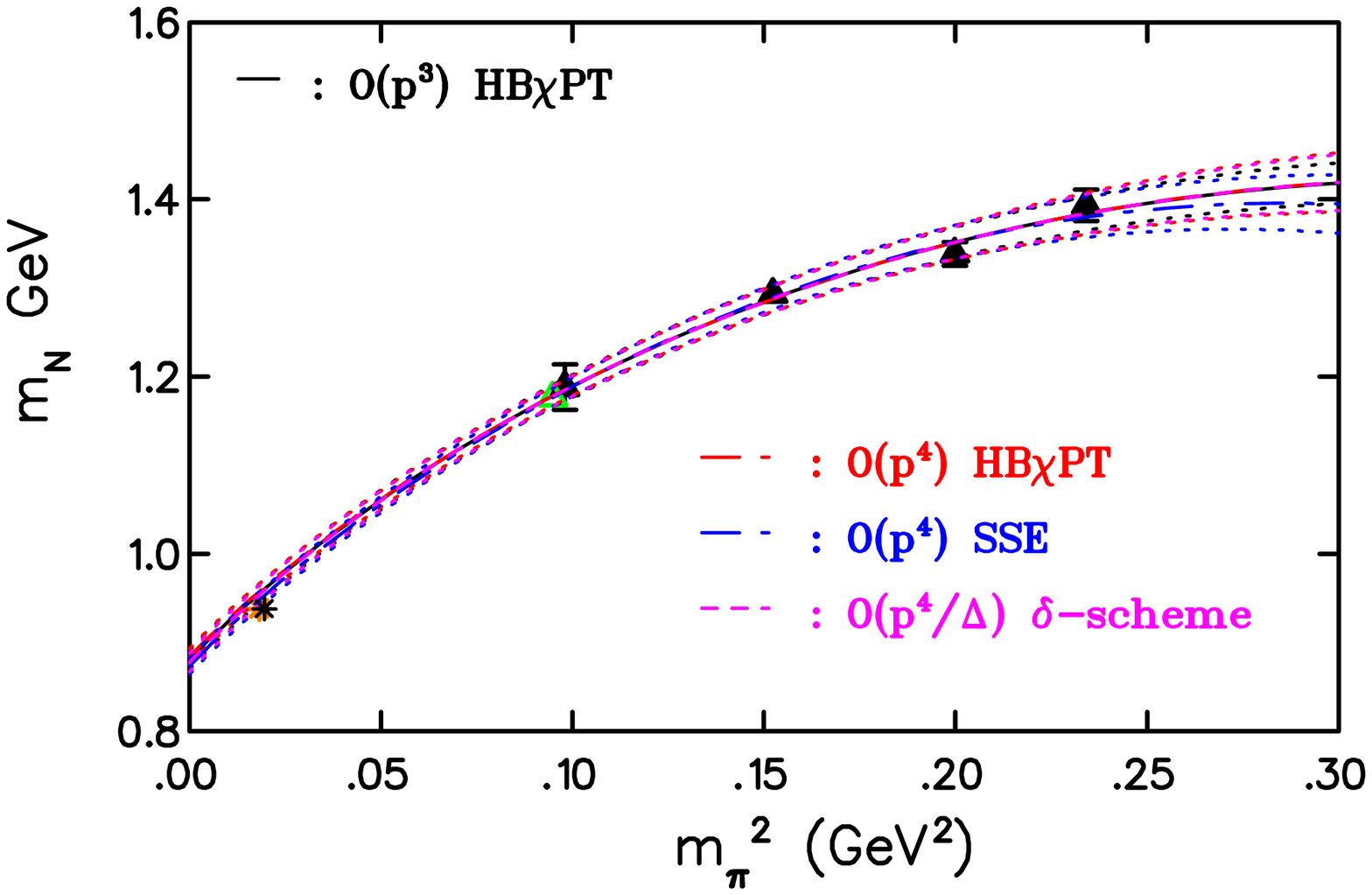}}}
\fi
\caption{Chiral fits to the nucleon mass using $a=0.0855$. The physical point
shown by the asterisk is not included in the fits.}
\label{fig:chiral fit mn}
\end{minipage}\hspace*{0.5cm}
\begin{minipage}[h]{7.cm}\vspace*{-1.2cm}
\ifpdf
{\mbox{\includegraphics[height=5.5cm,width=7.2cm]{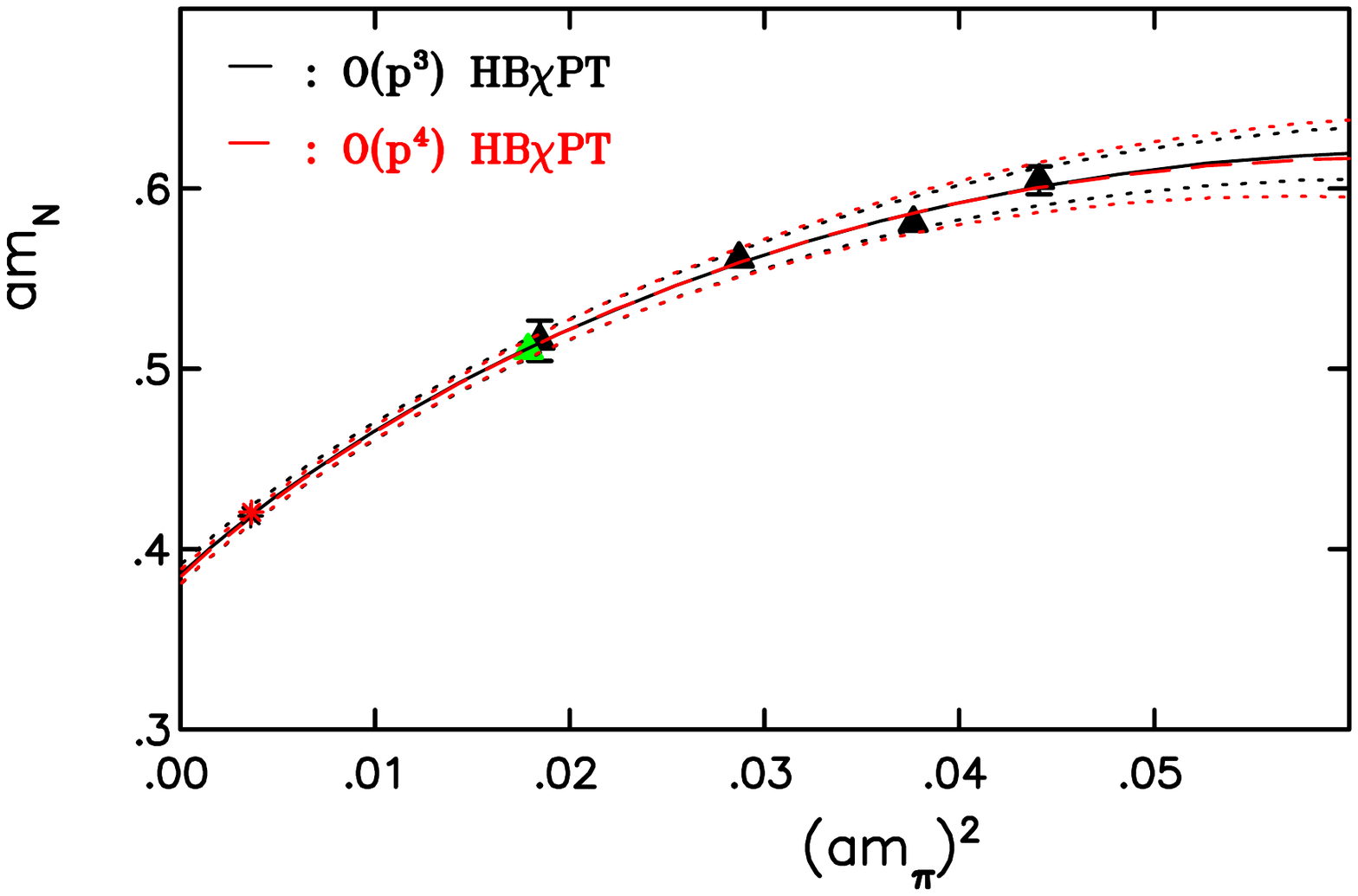}}}
\else
{\mbox{\includegraphics[height=5.5cm,width=7.2cm]{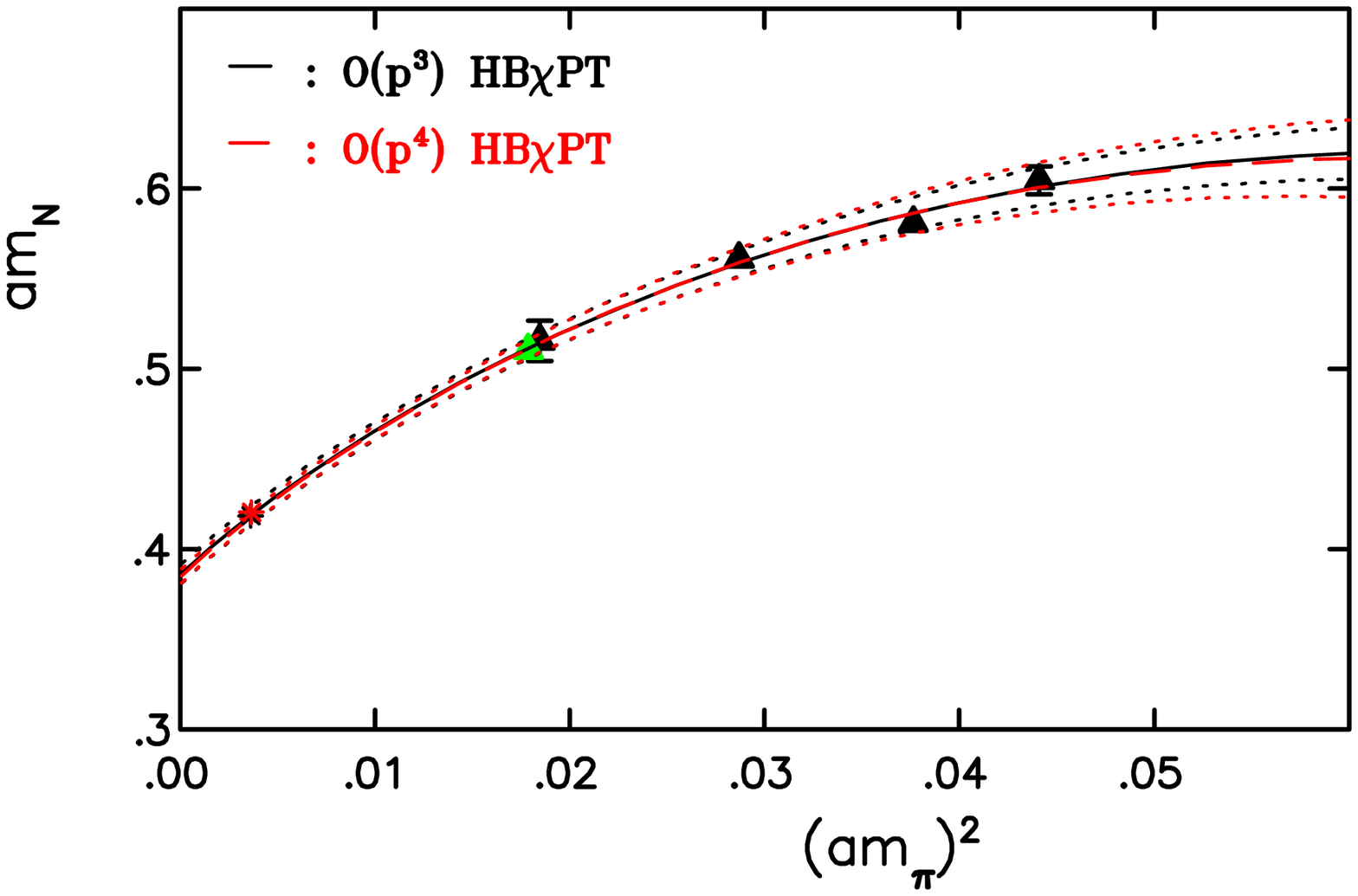}}}
\fi
\caption{Chiral fits using HB$\chi$PT to ${\cal O}(p^3)$
and ${\cal O}(p^4)$ determining $a$ from the nucleon mass.}
\label{fig:a from mn}
\vspace*{-1cm}
\end{minipage}
\end{figure}

Having checked  that $\Delta m_N$ at the smallest pion
mass is consistent with zero 
within our statistical errors and that  cut-off effects
are small,
we  use, in what follows, continuum chiral perturbation theory 
in an infinite volume to perform the chiral extrapolation to the
physical point.
The  leading one-loop result in heavy baryon chiral perturbation theory
(HB$\chi$PT)~\cite{HBchiPT} is well known:
\be 
m_N = m_N^0-4 c_1m_\pi^2- \frac{3g_A^2}{32\pi f_\pi^2}\> m_\pi^3
\label{mn to p3}
\ee
with $m_N^0$, the nucleon mass at the chiral limit, and $c_1$ treated as
fit parameters.  We find that this ${\cal O}(p^3)$ result
provides a very good fit to our lattice data at $\beta=3.9$, yielding
 $m_N^0= 0.875(10)$~GeV and 
$c_1=-1.23(2)$~GeV$^{-1}$ with
$\chi/{\rm d. o. f.}=0.2$. In this 
determination we use $a=0.0855$ and
results obtained on both lattice volumes. The value 
extracted for $c_1$ can be compared to the
value $c_1=-0.9\pm 0.5$~GeV$^{-1}$ extracted from various partial wave
analyses of elastic $\pi N$  scattering data
for the  $\pi N$-sigma term.
 We would like to stress that, despite the fact that the physical point is  
not included in the fit 
as customary done in other chiral extrapolations of
lattice data, the nucleon mass that we find at the physical pion mass is 
0.955(10)~GeV. Given that the error is only statistical, the fact that
this value  is so close to the
experimental value is  very satisfactory.
Chiral corrections to the nucleon mass are known to 
 ${\cal O}(p^4)$ within several
expansion schemes.
In this work to ${\cal O}(p^4)$   we use  the results
obtained in HB$\chi$PT with dimensional 
regularization~\cite{Bernard} and
in the so called small scale expansion (SSE)~\cite{Procura}. 
 HB$\chi$PT with dimensional 
regularization is in agreement with
covariant baryon $\chi$PT with infrared regularization
up to a recoil term
that is of no numerical significance. 
 In SSE the
$\Delta$-degrees of freedom are explicitly included in covariant baryon 
$\chi$PT 
by introducing as an additional counting parameter the
$\Delta$-nucleon mass
splitting, $\Delta\equiv m_\Delta-m_N$,  taking
${\cal O}(\Delta/m_N)\sim {\cal O}(m_\pi/m_N)$. 
A different counting scheme, 
known as $\delta$-scheme, takes
$\Delta/m_N\sim {\cal O}(\delta)$ and
$m_\pi/m_N\sim{\cal O}(\delta^2)$~\cite{Marc}.
 Using the $\delta-$scheme in a covariant chiral expansion to order $p^3$,
 $p^4/\Delta$
 one obtains an expansion that has a similar form 
for the nucleon and $\Delta$ mass.
Here we use the variant of the $\delta$-scheme that includes the 
$\pi\Delta$-loop and adds the fourth order term  $c_2 m_\pi^4$ 
as an estimate of higher order effects, since the complete fourth order
result is not available.
The parameter $c_2$ is to be determined from  the lattice data.
 The fits using these different formulations are shown in 
Fig.~\ref{fig:chiral fit mn}.
All formulations provide a good description of the lattice results
and yield a nucleon mass at the physical point that is close to 
the experimental value. The physical nucleon mass  is 
 not including in the fits.
We can use these chiral expansions 
 to fix the lattice spacing using the nucleon mass at the physical point and
compare with the value determined from the pion sector.
The results of the fits in HB$\chi$PT to ${\cal O}(p^3)$ and ${\cal O}(p^4)$ 
are shown in Fig.~\ref{fig:a from mn}.
Using the leading one-loop result we find  
$a=0.0879(12)$~fm,
whereas to ${\cal O}(p^4)$ 
 we obtain $a=0.0883(9)$~fm. 
Both SSE
and  the $\delta-$scheme,  which
include explicitly $\Delta$-degrees of  freedom,
 yield values that are  consistent with those obtained
in  HB$\chi$PT.
The variation in the value of $a$ in the
different chiral extrapolation schemes gives an estimation
of the systematic error involved in the chiral extrapolation.
A proper determination of the systematic error is in progress.

\begin{figure}[h]
\begin{minipage}[h]{7.7cm}\hspace*{-0.8cm}
\ifpdf
{\mbox{\includegraphics[height=5cm,width=7.cm]{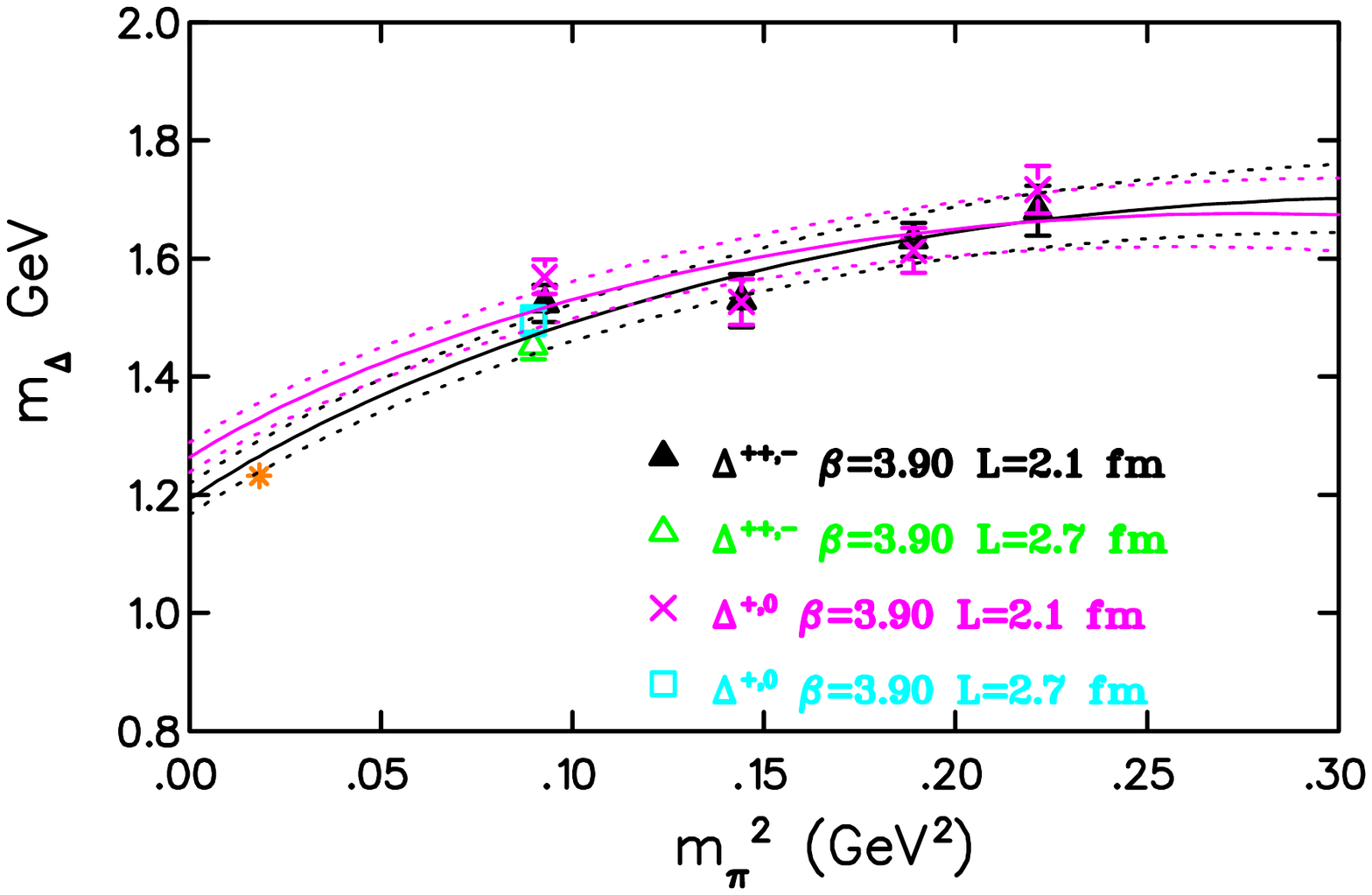}}}
\else
{\mbox{\includegraphics[height=5cm,width=7.cm]{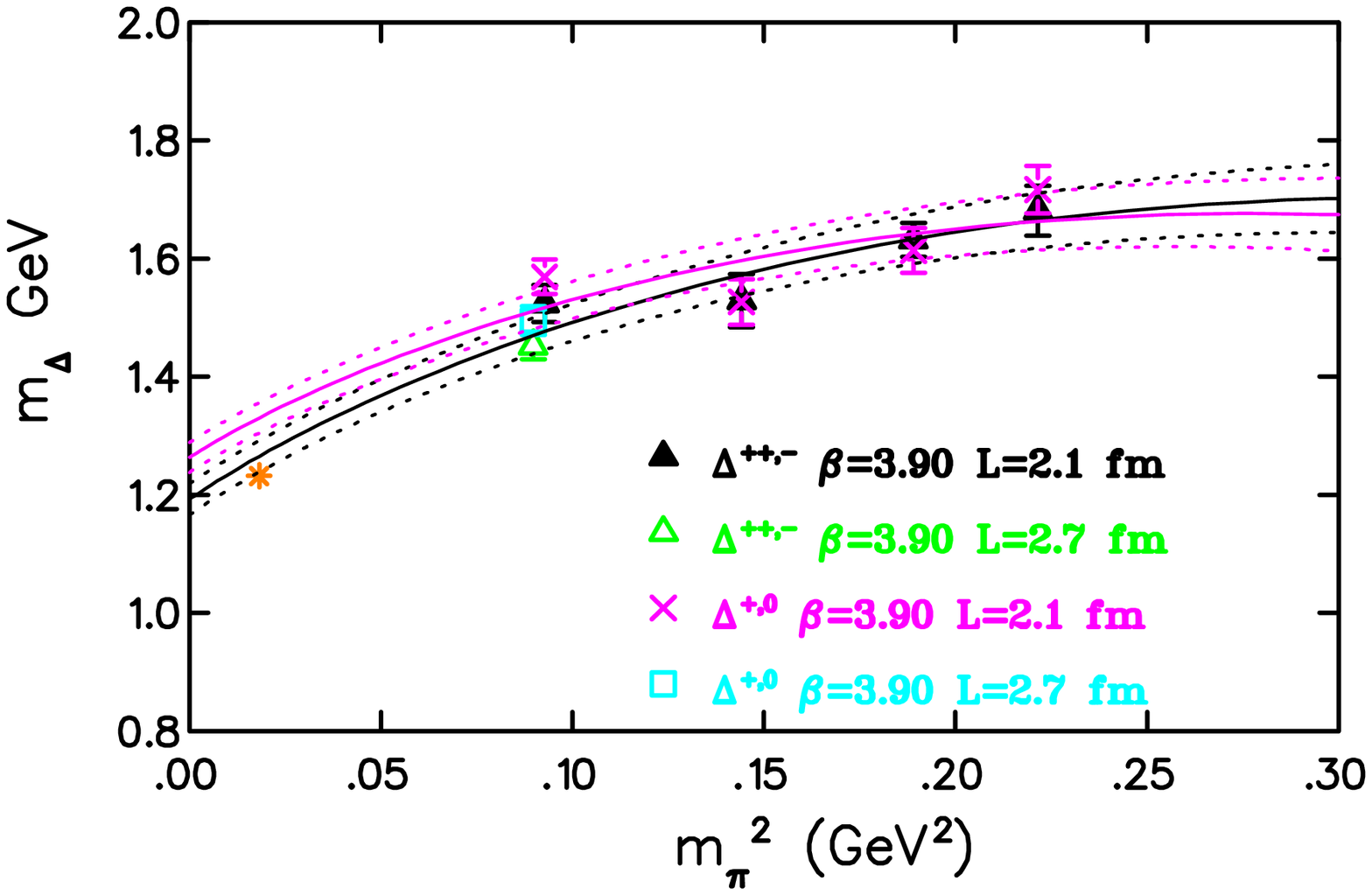}}}
\fi
\end{minipage}
\begin{minipage}[h]{7cm}\vspace*{-1.5cm}
%\item Use third order result:
%\be
% m_\Delta=m_0-4 c_1m_\pi^2- \frac{3g_A^2}{32\pi f_\pi^2}\> m_\pi^3
%\ee
%\small
\vspace*{0.8cm}
Lattice data on the $\Delta^{++,-}$ mass, chirally extrapolated using 
the  ${\cal O}(p^3)$-result in HB$\chi$PT, yield, at the physical point,
$m_{\Delta^{++,-}}=1.265(26)$~GeV, consistent with the resonant $\Delta$ mass.
A similar chiral fit to the $\Delta^{+,0}$ mass
yields a curve that lies above the physical point but
with an overall statistical error band 
that overlaps  the one obtained from the chiral fit to the
$\Delta^{++,-}$ mass.
\end{minipage}
\caption{Chiral fits to the $\Delta^{++,-}$ and $\Delta^{+,0}$ mass 
 using HB$\chi$PT to ${\cal O}(p^3)$
with $a$ set from the nucleon mass.}
\label{fig:chiral fits md}
\end{figure}

The leading one-loop HB$\chi$PT result  in the case of the $\Delta$ mass
has the same form as that for the nucleon mass given in 
Eq.~(\ref{mn to p3}) with $M_N^0 \rightarrow M_\Delta^0$
and $c_1 \rightarrow c_{1\Delta}$.
 Assuming SU(6) symmetry,
the one-loop contribution 
has the same numerical value  as in the nucleon case. 
It is useful to chirally extrapolate the $\Delta$ mass
to see how close current results are to $\Delta(1232)$ taking
$a=0.0879$ as determined from the nucleon mass. 
In Fig.~\ref{fig:chiral fits md} we show the resulting fit.

\newpage

\vspace*{-0.6cm}

\section{Conclusions}
We have shown that
twisted mass QCD yields  accurate results on the nucleon
mass close to the chiral regime.
The quality of the results for pion masses in the range of  300-500 MeV
allows a chiral extrapolation using heavy baryon
 chiral perturbation theory to
 ${\cal O}(p^3)$. The nucleon mass at the physical point
provides a good physical quantity for setting the scale.
 Using the leading one-loop
result in HB$\chi$PT we find  $a(\beta=3.9)=0.0879(12)$~fm.
Comparing this value to the results
obtained using 
 higher order terms in the chiral expansion, gives  a first
 estimate of the systematic uncertainty due to the
chiral extrapolation, that is of the  same order of  magnitude
as the statistical error.
Within this estimated uncertainty of the chiral extrapolation,
the value we find for $a(\beta=3.9)$ at leading order
in HB$\chi$PT is consistent with
the value  determined from $f_\pi$.
The mass splitting in the $\Delta$ isospin multiplets 
calculated with two lattice spacings on two volumes is consistent
with zero 
 showing that isospin breaking effects are not severe in this channel.

\end{document}